\documentclass[12pt]{article}
\usepackage{epsfig}
\usepackage{graphicx}

\usepackage{latexsym}
\usepackage{amsmath}
\usepackage{amssymb}
\usepackage{relsize}
\usepackage{geometry}
\usepackage{showlabels}
\def\beq{\begin{equation}}
\def\eeq{\end{equation}}
\def\beqn{\begin{eqnarray}}
\def\eeqn{\end{eqnarray}}

\def\slashed#1{\setbox0=\hbox{$#1$}             % set a box for #1
   \dimen0=\wd0                                 % and get its size
   \setbox1=\hbox{/} \dimen1=\wd1               % get size of /
   \ifdim\dimen0>\dimen1                        % #1 is bigger
      \rlap{\hbox to \dimen0{\hfil/\hfil}}      % so center / in box
      #1                                        % and print #1
   \else                                        % / is bigger
      \rlap{\hbox to \dimen1{\hfil$#1$\hfil}}   % so center #1
      /                                         % and print /
   \fi}                                        %

\begin{document}

%%%%%%%%%%%%%%%%%%%%%%%%%%%%%%%%

\begin{titlepage}

\begin{flushright}
ITEP-TH-38/11
\end{flushright}

\vspace{1cm}

\begin{center}
{  \Large \bf  SQCD, Superconducting Gaps and Cyclic RG Flows}
\end{center}
\vspace{1mm}
\centerline{\Large   A. Gorsky }

\vskip 0.5cm
\centerline{ {\em Institute of Theoretical and
Experimental Physics, Moscow 117259, Russia}}

%\begin{center}
\vspace {1.0 cm}
\centerline{ \Large \bf Abstract}

\vspace {5mm}
We consider the relation between the
$\Omega$  deformed  $N=2$ SQCD   with the single deformation parameter  $\epsilon$ and
integrable models of the BCS-like superconductivity. It is argued that  the
vortex string worldsheet theory
is related to the Russian Doll(RD) model of the truncated BCS superconductivity. We  argue
that the $\Omega$ deformed gauge theory manifests the interesting cyclic RG behavior
with the period of the RG cycle proportional to $\epsilon^{-1}$. The deformed gauge
theory can develop several  non-perturbative scales.
We conjecture on the monopole bound state interpretation
of the Efimov tower.

%\end{center}

\end{titlepage}

\section{Introduction}

A confinement of  electric degrees of freedom  implies a  condensation
of some magnetic degrees of freedom. They
certainly emerge from  D-branes however the precise pattern of the D-brane
condensation responsible for confinement in the asymptotically free gauge theories
is still unclear. One could imagine the condensation of the monopoles, strings,
domain walls or  complicated webs designed from these ingredients. Therefore
it is useful to look at the situations where the gap generation could be understood
from the first principles. The simplified models of superconductivity provide the
proper background to analyze the different aspects of these issues.

The well-known Seiberg-Witten solution \cite{sw} yields the pattern for
the monopole condensation in the perturbed $N=2$ SUSY YM theories with the different matter
content. It was argued that the  low-energy effective action
and the spectrum of the stable BPS states are governed by the classical
finite-dimensional integrable system \cite{gkmmm}.
In particular the theory which we shall mainly  focus on, namely  $N_f=2N_c$ SQCD,
is governed by the twisted XXX spin chain \cite{ggm1}. The vacua in the deformed
theory correspond to the classical equilibrium states
in the corresponding integrable system.
The relation between the finite dimensional integrable systems and
supersymmetric YM theories has been
generalized to the quantum case \cite{ns}. The parameter of the
$\Omega$ deformation plays the role of the Planck constant and the Bethe
anzatz equation in the finite dimensional integrable system corresponds
to the extremization of the superpotential in the deformed gauge  theory.

On the other hand the finite-dimensional integrable systems in some
cases provide the exact solution to the truncated BCS - like models of superconductivity.
The generic arguments explaining the relation between the integrable systems
and the superconducting models go as follows. Consider the fermionic
system with a kind of attractive four-fermion interaction. It can be derived upon
the integration over the "phonon"-like degrees of freedom.
The system  develops
the fermionic BCS condensate of the Cooper pairs which
is generically  inhomogeneous. It is
assumed that
the single fermion or Cooper pair propagating on the top of the condensate
should not ruin it. This is  a kind of consistency condition for the
combined system " inhomogeneous condensate + excitation".
It is this consistency condition that provides some integrable system behind the scene.
The space-time dependence of the condensate is governed by the integrable
equation itself while the equation for the one-particle fermion wave function in the condensate
background is interpreted as the Lax equation for the corresponding dynamical system.
The condensate can be represented via the continuous gap function
or some lattice of Abrikosov strings for the type II superconductivity.
In the first case one has the continuous integrable system of KdV type
while in the second case the discrete systems of Toda type are  involved.

The analogy with the Peierls superconductivity has been developed
in \cite{gorpei} for the pure perturbed $N=2$  SYM theory
whose low-energy sector  is described by the
periodic Toda chain \cite{gkmmm}.  The Pieirls model involves the
fluctuating  one-dimensional
crystal and fermions propagating in such background  \cite{revpei}. The Toda
system  governs the interaction between phonons
induced by fermions while the Lax equation describes the propagation
of the fermion in the  background of the emerging condensate. The Seiberg-Witten curve
is identified with the dispersion law for the fermions
and the nonperturbative $\Lambda_{QCD}$ scale in SYM corresponds to
the superconducting gap in the Peierls model.

The Toda system emerges as  the particular limit of the XXX spin chain when
the Casimirs of the representations  at each site of the chain tend to infinity. Hence we could
expect some superconducting model behind the SQCD as well
generalizing the arguments from \cite{gorpei}. It turns out that
such model exists indeed and we shall describe it in the paper.
The
Richardson model  \cite{richardson} and the RD model  \cite{rd} are proper
generalizations
for pure SQCD and $\epsilon$-deformed SQCD correspondingly.
The Richardson model can be also considered as the XY model in the
inhomogeneous external field.

On the other hand   the Richardson model links up with the
generalized Gaudin system \cite{richsolution} and  the  RD model with the twisted
inhomogeneous XXX spin chain correspondingly \cite{rdsolution}. It turns out that the relation between
the quantum integrable systems and the vacua of the
supersymmetric gauge theories \cite{ns} is quite convenient
to clarify the underlying physical degrees of freedom.
The Bethe anzatz equation describing the spectrum of the RD model
coincides with the equation describing the ground
state in the worldsheet theory of several nonabelian strings.
These semilocal nonabelian strings  with  rich worldsheet theory
\cite{book,eto}.
are the classical solution
to the equations of motion in the SQCD .

To some extend  the integrability plays the role of consistency condition
 in  more general setting. Recall that the most familiar
example of such phenomenon is provided by the strings propagating
in the external metric or the gauge field. It is well-known  that
the condition of the vanishing of Fradkin-Tseytlin  beta-function of
the string worldsheet theory
corresponds to the classical equations of motion
for background fields. Speaking differently  this is the consistency condition
for the propagation of the string in its own condensate.

In the
problem under consideration the situation is more subtle since
the worldsheet theory of the nonabelian string  is not conformal.
Hence the consistency condition
is more involved and  the quantum properties of the bulk fields have to be matched in this case. The
interplay between the $D=4$ bulk theory and the $D=2$ worldsheet theory
on the nonabelian string has been examined in \cite{sy,doreyold} where it was
argued that the $\beta$-functions and the spectrum of the BPS states
match. In this paper we shall argue that the additional ingredient
can be  added to the generic matching condition that is
the system of superconducting fermions. The corresponding Richardson
of RD models are conjectured to describe the magnetic degrees of freedom
and the consistency
condition implies that their dispersion law fits with the spectral
curve of the proper  integrable system. These integrable spin chain models
correspond to the dynamics of the nonabelian strings per se.

Due to the mapping of  $\epsilon$ -deformed  SQCD  into the RD model it is possible
to gain  interesting observations from the known results concerning the superconducting model.
The most unusual feature
of RD model is the cyclic RG behavior yielding  the
multiple condensates instead of the single one \cite{rd}.
The period of the RG cycle is determined by the RG  invariant
and the spectrum of the model is reshuffled upon each cycle.
There are several systems enjoying  cyclic RG
flows both in the quantum mechanics \cite{wilson} and in the field theory with
the several couplings \cite{ft}. In particular such behavior has been found
in the sin-Gordon model analytically continued in the coupling
constant \cite{ft}. In that example the cyclic RG behavior amounts
to the set of the unusual resonances with the Regge-type
stringy spectrum.  In general the origin of the cyclic RG
behavior is some resonance-like behavior in two-body system
which amounts into the hierarchy of the Efimov-like states.
The most surprising aspect of the cyclic RG is its sensitivity
to the UV scale of the theory under consideration. Note that
is was shown recently that the cyclic RG flow is consistent with
the c- and a-theorems \cite{curt}.

 The
RG step in RD model corresponds to the change of the  XXX spin chain
length.  Being translated to the conformal gauge theory side it corresponds
to the decoupling of  two flavors sending the corresponding
mass to infinity with the simultaneous
change $N \rightarrow (N-1)$. The theory  remains conformal
with the different rank of the gauge group. However
in the
deformed theory we have additional dimensionless parameter which is the ratio
$\frac{\epsilon}{m}$ hence
decoupling of the heavy flavor is described via  two-coupling RG flow.
Using the known results concerning  RD model  we shall argue that
decoupling of a heavy  flavor occurs in a cyclic manner and
the period of a cycle is proportional to the inverse of the
deformation parameter therefore in the undeformed case the period is infinite.

The remarkable AGT relation \cite{agt} makes manifest
the connection between the SU(2) Nekrasov partition function
of the deformed SYM theory  and the
conformal blocks in the Liouville theory \cite{agt}. The interplay between
the four-dimensional and two-dimensional theories reflects the
complimentary viewpoints of  observers at the  M5 brane  worldvolume 
where the SYM lives on.  For the higher rank group the
Liouville theory gets substituted by the 2d Toda theory \cite{wyllard}.
The nonabelian strings with large tension  in the gauge theory are identified
as the surface operators which on the other hand are represented as the
proper vertex operators  in the Liouville-Toda  theory \cite{surface}.
In particular the D2 branes representing the surface operators
in Liouville-Toda theory yield the degrees of freedom for the Toda-Calogero type
integrable models.

Hence we could look at the possible interpretation of the cyclic
RG flows for the conformal blocks in the Liouville/Toda theory and
the possible interpretation of the superconducting models on the
gravity side of the AGT correspondence. The key point is the interpretation
of the fermions and their Cooper pairs from the first principles. We shall make
some conjectures however the complete analysis is still to be done.

The paper is organized as follows. In Section 2 we briefly review
the Peierls model while in Section 3 the Richardson model and the
RD model are considered.  In Section 4 we comment on the generic
properties of the cyclic RG flows. In Section 5
the mapping to
the vortex nonabelian states in the SQCD is considered. In Section 6 we
conjecture on the
interpretation of the fermions in the superconducting model
as the monopoles in the Higgs branch.
The main observations of the paper and the list of the
open questions can be found in the last Section.

\section{The Peierls model}

In this section we briefly review  the main facts concerning the Peierls model relevant
for the further consideration. Originally it was formulated to describe the
selfconsistent behavior of 1d fermions interacting with the fluctuating
lattice and has been  applied to the description of the   1d superconductivity \cite{revpei}. The Coulomb
interaction between the fermions is neglected, the fermions are assumed to be
in the external field determined by the lattice degrees of freedom while
the lattice dynamics
itself gets modified by the fermions. In what follows we will
mention the integrable structure both in  the continuum \cite{belokolos} and discrete Peierls models \cite{revpei}.
Hamiltonian density for the  simplest continuum model looks as follows
\beq
H_{con}=\Psi^{+}\sigma_{3}\partial_{x}\Psi +\Psi^{+}(\sigma_{-}\Delta(x)^{*}-
\sigma_{+}\Delta(x))\Psi +\Delta(x)^{2}
\eeq
where $\Delta(x) $ represents the lattice potential. The saddle point solution
for $\Delta(x)$ provides the four-fermion interaction Hamiltonian
induced by the phonon exchange. To some extend the lattice
potential $\Delta(x)$ can be interpreted as the inhomogeneous fermionic
condensate.

For the discrete
version one has   \cite{revpei}
\beq
H_{dis}=\sum_{n}
(\Psi_{n}^{+}v_{n}\Psi_{n}+\Psi_{n}^{+}c_{n}\Psi_{n+1}+
\Psi_{n}^{+}c_{n-1}\Psi_{n-1}) +\sum_{i}\kappa_{i}I_{i}
\eeq
\beq
c_{n}=exp(x_{n+1}-x_{n}),\qquad
I_{0}N=\sum_{n} lnc_{n}, \qquad I_{2}N=\sum_{n}(c_{n}^{2}+v_{n}^{2})
\eeq
where $x_{n}, v_n$ are the lattice coordinate and momentum
while $I_n$ are identified as the Toda chain Hamiltonians.

To identify  the ground state of the model we  minimize the Hamiltonian with respect to
the fermionic and lattice variables. The variation over the fermionic
variables yields the Lax equation for the Toda chain
\beq
c_{n}\Psi_{n+i}+c_{n-1}\Psi_{n-1}+v_{n}\Psi_{n}=E\Psi_{n}
\eeq
Variation over the lattice degrees of freedom gives rise to the system of the
finite number of algebraic equations describing the Riemann surface.
In the Peierls model we assume the  periodicity
of the fermionic wave function on the lattice
\beq
\Psi_{n+N}(E)=e^{iNp(E)}\Psi_{n}(E) .
\eeq
The solution is described in terms of  the
hyperelliptic Riemann surface
\beq
y+y^{-1}=P_{N}(E)
\eeq
where $P_N(E)$ is the polynomial.
This Riemann surface  plays the double role. First, its Jacobian
is identified  as the complex
Liouville torus for the periodic Toda chain
describing the dynamics of phonons. On the other hand it is
nothing but the Fermi surface for the fermions. The moduli of this surface
are fixed by the minimization equations
in terms of the physical parameters like the fermionic density $\rho=\frac{N_e}{N}$.
The key feature of the solution
is the appearance  of the fermionic mass gap  which
is analogue of the $\Lambda_{QCD}$.
Fermionic wave function is uniquely defined on this
surface and the number of its zeros coincides with the genus of the curve.
It was proved  that only one gap vacuum configuration
is stable, therefore the generic polynomial degenerates.
The very similar one gap vacuum configuration emerges
in the Gross-Neveu model \cite{dunne}.

The spectrum of the excitations  above the ground state involves
fermionic and bosonic quasiparticles. The  phonon and
the charge density wave represent the bosonic excitations \cite{belokolos,dk}.
The fermionic excitations  strongly depend on the fermionic
density $\rho$. At large $\rho$   the polyaron type state is
localized at the lattice configuration
\beq
\Delta(x)=const-\frac{2\chi}{ch^{2}(\sqrt{\chi}x)}
\eeq
In the opposite limit at  small density one has
the delocalized fermionic state
and the gap potential
\beq
\Delta(x)=const + \chi cos(2\sqrt{\chi}x+\phi)
\eeq
where $\chi$ is some constant.
Exact solution  provides the
temperature dependence of the mass gap. It was shown \cite{belokolos}
that the fermion mass gap
gets renormalized and disappears at some critical value
$T_{c}$. Being translated into the form of the dispersion law it tells us that
the Riemann surface degenerates to a sphere above the phase transition point.

%\section{Integrable structure of SUSY YM theories and the dispersion laws}

Let us  compare the data governing the vacuum structure of the pure $N=2$
SUSY YM theory and the one from the Peierls model. Low energy effective action
in SYM theory is fixed by the Riemann surface and holomorphic differential
defined on it  \cite{sw}. Prepotential $\cal{F}$ can be derived from the relations
\beq
a_{D_{i}}=\frac{d\cal{F}}{da_{i}}\qquad
a_{i}= \oint_{A_{i}}\lambda \qquad
a_{D_{i}}=\oint_{B_{i}}\lambda \qquad  \lambda=dS=Edp
\eeq
where $A_i$ and $B_i$ are the cycles on the Seiberg-Witten spectral curve.
The number of cites in the Peierls model corresponds to the rank of the gauge group while  the
total length can be identified with the coupling constant
$\tau_{0}=\frac{4\pi i}{g^{2}}+\frac{\theta}{2\pi}$
taken at the UV scale. The
density of the fermions turns out to be related to the renormalized coupling constant
in the gauge theory \cite{gorpei}. The equations selecting the ground state of the Peierls model
correspond to the selection of the point at the Coulomb branch of the moduli space.

It what follows we shall generalize the relation between the  low-energy
sector of SYM theory and models of superconductivity for the theories with
matter in the fundamental representation.

\section{Truncated models of BCS superconductivity}
\subsection{Richardson model versus Gaudin with irregular singularities}

Let us recall the truncated BCS-like Richardson model of superconductivity \cite{richardson}
with some number
of doubly degenerated fermionic levels with the energies $\epsilon_{j\sigma} \dots j=1\dots N$.
It describes the system
of a  fixed number of the Cooper pairs.
It is assumed that  several energy levels are populated by  Cooper pairs while
levels with the single fermions are blocked.
The Hamiltonian reads
as
\beq
H_{BCS}= \sum_{j,\sigma= \pm}^N \epsilon_{j\sigma} c^{+}_{j\sigma}c_{j\sigma} -
G\sum_{jk}c^{+}_{j+}c^{+}_{j -}c_{k -}c_{jk+}
\eeq
where $c^{+}_{j\sigma}$ are the fermion operators and $G$ is the
coupling constant providing the attraction leading to the formation
of the Cooper pairs. It terms of the hard-core boson operators
it reads as
\beq
H_{BCS}=\sum_j\epsilon_j b^{+}_jb_j - G \sum_{jk}b^{+}_jb_k
\eeq
where
\beq
[b^{+}_jb_k]= \delta_{jk}(2N_j-1), \qquad b_j=c_{j -}c_{j +},\qquad N_j= b^{+}_jb_j
\eeq

The eigenfunctions of the Hamiltonian can be written as
\beq
|M>=\prod_i^M B_i(E_i)|vac>, \qquad B_i= \sum _j^N \frac{1}{\epsilon_j- E_i} b^{+}_j
\eeq
provided the Bethe anzatz equations are fulfilled
\beq
\label{BA}
G^{-1}= - \sum _j^N \frac{2}{\epsilon_j- E_i} + \sum _j^M \frac{1}{E_j- E_i}
\eeq
The energy of the corresponding states reads as
\beq
E(M)= \sum_i E_i
\eeq

It was shown in \cite{richsolution} that the Richardson
model is exactly solvable and closely related to the particular generalization
of the Gaudin model known as the model with  irregular singularities
\cite{irregular}. To describe this relation it is convenient
to introduce the so-called pseudospin $sl(2)$ algebra in
terms of the creation-annihilation
operators for the Cooper pairs
\beq
t^{-}=b \qquad t^{+}=b^{+} \qquad t^{0}=N-1/2
\eeq

The Richardson Hamiltonian commutes with
the set of operators $R_j$
\beq
R_i= -t^0_i -2G\sum^N_{i\neq j}\frac{t_i t_j}{\epsilon_i -\epsilon_j}
\eeq
which are identified as the Gaudin Hamiltonians.
\beq
[H_{BCS},R_j]=[R_i,R_j]=0
\eeq
Moreover the Richardson Hamiltonian itself
can be expressed in terms of the operators $R_i$ as
\beq
H_{BCS}= \sum_j \epsilon_i R_i + G(\sum R_i)^2 + const
\eeq

The number of the fermionic levels N coincides
with the number
of sites in the Gaudin model and the coupling constant in the
Richardson Hamiltonian corresponds to the "twisted boundary
condition" in the Gaudin model. The Bethe anzatz equations
for the Richardson model (\ref{BA}) exactly coincides with
the ones for the generalized Gaudin model.
Note that the Gaudin model
is the Hitchin system on the marked sphere which
plays the role of the Fermi surface for the fermions very much
in the same manner as we have seen in the Peierls model.
 It was argued in \cite{rd} that the
Bethe roots corresponds to the excited Cooper pairs that
is natural to think about the solution to the Baxter
equation as the wave function of the condensate. In terms
of the conformal field theory Cooper pairs correspond
to the screening operators \cite{sierraconf}.

For the nontrivial degeneracies of the energy levels $d_j$
the BA equations read as
\beq
G^{-1}= - \sum _j^N \frac{d_j}{\epsilon_j- E_i} + \sum _j^M \frac{2}{E_j- E_i}
\eeq

\subsection{Russian Doll model of superconductivity and  twisted XXX spin chains}

The important generalization of the Richardson model
describing the reduced BCS superconductivity as well
is the so-called RD model \cite{rd}. It involves the
additional dimensionless parameter $\alpha$ and the RD Hamiltonian reads as
\beq
H_{RD}= 2\sum_i^N (\epsilon_i- G)N_i -\bar{G}\sum_{j<k}
(e^{i\alpha} b^{+}_k b_j +e^{-i\alpha} b^{+}_jb_k)
\eeq
with two dimensionful parameters $G, \eta$
and $\bar{G} =\sqrt{G^2 +\eta^2}$. In terms of these variables the
dimensionless parameter
$\alpha$  has the following form
\beq
\alpha= arctanh(\frac{\eta}{G})
\eeq
It is  also useful to consider   two dimensionless parameters $g,\theta$
defined as $G=gd$ and $\eta =\theta d$ where $d$ is the level spacing.
The RD model reduces to the Richardson model in the limit $\eta\rightarrow 0$.

The RD model turns out to be integrable as well. Now instead
of the Gaudin model the proper counterpart
is  the generic quantum twisted XXX spin chain \cite{rdsolution}
The transfer matrix of such spin chain
model $t(u)$ commutes with the $H_{RD}$ which itself can be
expressed in terms of the spin chain Hamiltonians.

The equation defining the spectrum of the RD model reads as
\beq
\exp(-2i\alpha)\prod^N \frac{E_j-\epsilon_k -i\eta/2}
 {E_j-\epsilon_k + i\eta/2}=
 \prod^M \frac{\epsilon_j-\epsilon_k -i\eta}
 {\epsilon_j-\epsilon_k + i\eta}
 \eeq
and coincides with the BA equations for the spin chain.
It reduces to the BA equation of the Richardson model
(\ref{BA})  in the limit
 $\eta \rightarrow 0$.

 The key feature of the RD model is the multiple
 solutions to the gap equation. The
 gaps are parameterized as follows
 \beq
 \Delta_n= \frac{\omega}{sinht_n}, \quad t_n= t_0+ \frac{\pi n }{\theta} \quad n=0,1 \dots
 \eeq
 where $t_0$ is solution to the following equation
  \beq
  tan(\theta t_0)= \frac{\theta}{g}\qquad 0<t_0< \frac{\pi}{\theta}
  \eeq
and $\omega =dN$ for  equal level spacing. This
behavior can be derived via the mean field approximation \cite{lrs}.
Different solutions to the gap equation  yield the
different  superconducting states.
In the limit $\theta \rightarrow 0$
the gaps $\Delta_{n>0} \rightarrow 0$ and
\beq
t_0=\frac{1}{g},\qquad \Delta_0= 2\omega e^{\frac{1}{g}}
\eeq
therefore the standard BCS
expression for the gap is recovered. At the weak coupling
limit the gaps behave as
\beq
\Delta_n \propto \Delta_0 e^{-\frac{n\pi}{\theta}}
\eeq
In terms of the solutions to the BA equations the
multiple gaps correspond to the choices of the
different branches of the logarithms.

If the degeneracy of the levels is $d_n$ then the
RD model gets modified a little bit and is related to the
higher spin XXX spin chain. The local spins $s_i$ are determined
by the corresponding higher pair degeneracy $d_i$ of the i-th level
\beq
s_i=d_i/2
\eeq
and the corresponding BA equations read as
\beq
\exp(-2i\alpha)\prod^N \frac{E_j-\epsilon_k -i\eta/2+ i\eta s_i}
 {E_j-\epsilon_k + i\eta/2 - i\eta s_i}=
 \prod^M \frac{\epsilon_j-\epsilon_k -i\eta}
 {\epsilon_j-\epsilon_k + i\eta}
 \eeq
When the local spin tends to infinity the twisted XXX chain degenerates
into the periodic Toda chain hence the large degeneracy of the
levels is necessary at the superconducting side. It is this limit
which corresponds to the Peierls model discussed earlier.

\section{Cyclic RG flows and RD model}
Let us explain the key points concerning the cyclic RG phenomena.
It was anticipated for a while however the first clear-cut example
has been elaborated only in \cite{wilson} for the finite-dimensional
system. The method developed in \cite{wilson} formulates the step in the RG flow
in the finite-dimensional system as removing the single highest energy level
with simultaneous renormalization of the couplings. This
approach has a lot in common with the renormalization procedure in the
matrix models considered in \cite{brezin}. The different approach to
RG in the finite-dimensional system concerns the introducing the
UV cutoff at the small scales and looking at the cutoff dependence. This approach
has been successfully applied to the rational Calogero model \cite{calogero}
and it was argued that  conformal quantum mechanics manifests the
cyclic RG which reflects a kind of "anomalous" violation of the conformal group
down to the discrete one.

In fact the important phenomena which has the cyclic RG origin has been discovered
long time ago by Efimov in the three-body system . His key observation
concerns the emergence of the specific bound states in the three-body system
involving two very different scales. It was argued that in such situation
when there are the two-body bound state near threshold there is the tower of the
bound states with the RD scaling behavior of  energies
\beq
E_{n+1}=e^sE_n
\eeq
The presence of
so-called Efimov states with the RD scaling is the common feature of
the models with cyclic RG phenomena. The spectrum is reorganized in the universal
way during the single cycle and the total number of Efimov states
is of order $log(\frac{r_{UV}}{r_{IR}})$. The review on the cyclic RG in the
finite-dimensional systems can be found in \cite{braaten}.

There are  examples of the cyclic RG flows in the 2D field theories
with the several couplings where the RG is formulated in the standard manner
as the dependence on the log of the renormalization scale \cite{ft}. It was argued
that generically there should be at least two couplings and usually
one of them does not run at all. In the field theory  one also  has the
RD scaling for the Efimov-like resonances which in some examples
manifest the Regge like structure. Moreover it was argued that the
S-matrix behaves universally under the cyclic RG flows. The total number
of Efimov states scales in the same manner as in the quantum mechanical case.

 The RD model of truncated superconductivity enjoys the cyclic RG behavior \cite{rd}.
 The RG flows can be treated as the integrating out the
 highest fermionic level with appropriate  scaling
 of the parameters using the procedure developed in
 \cite{bhk,wilson}. The RG equations read as
 \beq
 g_{N-1}=g_N + \frac{1}{N}(g_N^2 + \theta^2),\qquad \theta_{N-1}=\theta_N
 \eeq
 At  large N limit the natural RG variable is
 identified with $log N$ and the solution to the
 RG equation is
 \beq
 g(s)=\theta tan(\theta s + tan^{-1} (\frac{g_0}{\theta}))
 \eeq
 Hence the running coupling is cyclic
 \beq
 g(s+ \lambda) =g(s),\qquad g(e^{-\lambda}N)=g(N)
 \eeq
 with the RG period
 \beq
 \lambda=\frac{\pi}{\theta}
 \eeq
and the total number of the independent gaps in the model is
\beq
N_{cond}\propto \frac{\theta}{\pi} log N
\eeq
The multiple gaps  are the manifestations
of the Efimov-like states. The sizes of the Cooper
pairs in the n-th  condensates also have the
RD scaling. The cyclic RG can be derived even for the single
Cooper pair.

What is going on with the spectrum of the model during the period? It was
shown in \cite{lrs} that it gets reorganized. The RG flows
possesses the discontinuities from $g=+\infty$ to $g=-\infty$
when a new cycle gets started. At each jump the lowest
condensate disappears from the spectrum
\beq
\Delta_{n+1}(g=+\infty)= \Delta_{n}(g=-\infty)
\eeq
indicating that (N+1)-th  state wave function
plays the role of N-th state wave function at the next cycle.

The same behavior can be derived from the BA equation \cite{lrs}.
To identify the multiple gaps it is necessary to remind
that the solutions to the BA equations are classified by the
integers $m_i, i= 1,\dots M$ parameterizing the branches of the
logarithms. If one assumes that $m_i=m$ for all Bethe roots
then this quantum number gets shifted by one
at each RG cycle  and was identified with the integer
parameterizing the solution to the gap equations At the
large N limit the BA equations of the RD model reduce
to the BA equation of the Richardson-Gaudin model
with the rescaled coupling
\beq
G_{m}^{-1}= \eta^{-1}({\alpha} + \pi m)
\eeq
which can be treated as the shifted boundary
condition in the generalized Gaudin model
parameterized by the integer.
Let us  emphasize that  the unusual cyclic RG behavior
in due to the presence of two couplings in the RD model.

\section{Nonabelian string versus BA equation}
\subsection{BA equations at the string worldvolume}
Let us consider the $\Omega$ deformed $N=2$ SQCD with $SU(L)$
gauge group, L fundamental hypermultiplets with masses $m_{f_i}$ and L antifundamental
hypermultiplets with masses $m_{afi}$.
In the NS limit with the single
deformation parameter $\epsilon$ the Coulumb branch of the vacuum manifold
parameterized by the vev of the adjoint scalar $a_i$
gets reduced to the set of points since the theory develops
the twisted superpotential \cite{ns} determined by
Nekrasov partition function.
The isolated vacua are determined by the equation
\beq
\frac{\partial {\cal W}(a_i)}{ \partial a_i} = n_i
\eeq
which yields
\beq
\vec{a}= \vec{m_{f}} -\vec{N}\epsilon
\eeq
where $\vec{n}=(n_1\dots n_L)$.
The integers can be attributed to the selection
of the branch of logarithm  in the twisted superpotential.

The bulk D=4 theory develops the Higgs branch where the
scalar components of the fundamental are condensed. The
Higgs branch admits the stable nonabelian string
solution with the rich worldsheet theory.
The physics of the non-abelian strings in the nondeformed case is described
in the reviews \cite{book}.

In the bulk theory in the NS limit
the worldsheet theory on the nonabelian string  involves the
L fundamental chiral multiplets with twisted masses $M_{F_i}$
and L antifundamental multiplets with twisted masses $M_{AF_i}$.
The additional chiral multiplet in the adjoint representation
gets non-vanishing mass $\epsilon$ due to the  the background
graviphoton field. It can be integrated out amounting
to the twisted superpotential for the scalars in the vector
multiplet. The minimization of the worldsheet theory at N
nonabelian strings yields the BA equations where Bethe roots
$\lambda_i$ correspond to the values of the scalars in the
vector multiplet \cite{ns,dorey,dorey2}. The solutions to the BA equations are parameterized
by the set of integers $\tilde{n_i}$ obeying the condition $N=\sum_{i=1}^L n_i$.

It turns out that  on-shall values of the twisted superpotentials
in the bulk theory and the worldvolume theory of N
nonabelian strings coincide
\beq
W^{4D}(a_l=m_l-n_l \epsilon) - W^{4D}(a_l=m_l- \epsilon) = W^{2D}({\tilde{n}_L})
\eeq
upon the following identifications
of parameters \cite{dorey}
\beq
 \vec{ M}_f= \vec{m}_f - 3/2 \epsilon,\qquad \vec{M}_{af}= \vec{m}_{af} -1/2 \epsilon
\eeq
The rank of the worldvolume theory $N$ which
corresponds to the number of nonabelian strings
is defined via the relation
\beq
N + L = \sum_{i} n_i, \qquad \tilde{n}_l= n_l-1
\eeq
The   modular  parameter in the worldsheet theory
\beq
\tau_{2D} = ir + \frac{\theta_{YM}}{2\\pi}
\eeq
where $r$ is the FI parameter  related to
 the $D=4$ coupling constant  \cite{sy}.
The $\theta$ term  penetrates the worldsheet theory from
the bulk one \cite{gsy}.
The modular parameters of the theories are related as
\beq
\tau_{2D} =\tau_{4D} +\frac{1}{2}(N+1)
\eeq

In the $D=4$ bulk theory the BA equations emerge in the
saddle point calculation of the instanton partition function \cite{dorey2}.
The number of the Bethe roots
coincides with the number of the nonabelian strings
which is in the perfect agreement with the interpretation
of the nonabelian strings as the excitations
above the root of the Higgs branch in the bulk theory.

The BA equations for the deformed SQCD exactly coincide
with the BA equations involved into the solution to
the RD model. The asymmetry parameter of the
RD model $\eta$ is identified with the deformation
parameter of the  $\Omega$ background
\beq
\eta= \epsilon
\eeq
The fermionic energies in the RD model
are identified with the masses of fundamentals
in $D=4$ bulk theory or the twisted masses in the
worldsheet theory
\beq
E_i= M_i
\eeq
The   modular  parameter in the worldsheet theory
 provides the
twisted boundary conditions in the BA equations
for the inhomogeneous XXX spin chain.

Geometrically the Bethe root  or vev of the scalars corresponds
to the position of the D2 brane representing
nonabelian string in the transverse coordinates. Note that
from the  spin chain viewpoint the Bethe roots
are zeros of the polynomial solution
\beq
Q(\lambda)= \prod (\lambda -\lambda_i)
\eeq
to the Baxter
equation for the twisted spin chain.

\subsection{Cyclic RG in the deformed gauge theory}

Let us turn to the important observation.
Since the BA equation for the deformed SQCD coincides
with the one for the twisted XXX spin chain we can
use the results concerning its  cyclic RG behavior.
First remind once again that there are two dimensionless couplings
in the $\Omega$ deformed SQCD required
for cyclicity. One of them is the conventional
complexified coupling in the bulk gauge theory
which plays the role of the twist in the spin chain.
The second dimensionless parameter in the simplest case can
be identified with the ratio
\beq
\theta= \frac{\epsilon}{\delta m}
\eeq
where $\delta m$ is the difference between the
masses of the fundamentals. We assume that the
masses are almost equidistant. It is useful
following \cite{lrs} to introduce the
second dimensionful parameter $G$ similar to the
$\Lambda_{QCD}$ scale
\beq
tan \tau = \frac{G}{\epsilon}
\eeq
This scale is evidently nonperturbative
with respect to the gauge coupling.

To describe the cyclic RG behavior in SQCD
let us identify the step of the RG flow
in terms of the finite-dimensional system.
As we have mentioned the step in the RD model
corresponds to the decoupling of the highest
energy level. Since
the energy level in RD model corresponds
to the mass of fundamental the RG step
corresponds to the decoupling of the
heavy flavor $N_F\rightarrow N_F-2$.
Simultaneously the rank of the group
is changed $N_c\rightarrow N_c -1$ and
the form of the BA equation holds the same.

In the non-deformed case this procedure
yields the RG flow without any cycles.
On the other hand if we just decouple
the heavy flavor without changing
the rank of the group
the nonperturbative scale emerges
in the asymptotically free theory.
\beq
\Lambda = M_{reg} \exp (\frac{2\pi}{\alpha(M_{reg})\beta_0})
\eeq
where $M_{reg}$ is the UV scale and $\beta_0$
is the coefficient of the $\beta$-function.
In the deformed case the situation is more
involved since the decoupling
of the heavy flavor is described by 
two-coupling RG equations. As we have seen before
the $\epsilon$ parameter is not deformed however
the  modular parameter enjoys the cyclic RG
solutions. Indeed the complexified gauge coupling
is expressed as a function of the scales via (\ref{coup})
hence the RG flow of G yields the RG flow of $\tau$.
Let us emphasize that we consider the conformal
theories and the cyclicity involves the rank of the
gauge group. The ciclicity of the RG flow breaks
the conformal group down to the discrete subgroup
selected by the deformation parameter.

As we have described above the key feature
is the emergence of the multiple Efimov-like
scales in the problem. In the RD model
these Efimov-like scales correspond to
the multiple gaps with the Efimov scaling.
In the deformed SQCD this scales emerge as
the multiple nonperturbative $\Lambda$-like scales
whose number
is defined by the ratio
\beq
N_{cond}\propto \frac{\epsilon}{\delta m}
\eeq
At  the weak coupling  the scales behave as
\beq
\Delta_n\propto \Delta_0 e^{-\frac{n\pi \delta m}{\epsilon}}
\eeq
which are certainly nonperturbative in the parameter of
the $\Omega$ background.

Our observations provide the qualitative picture behind
the RG flows in the deformed QCD and more detailed analysis
is required. In particular we have considered the situation
with the single massive parameter only. In the generic situation
we have the set of the dimensionless parameters
\beq
\theta_i= \frac{m_i}{\epsilon}
\eeq
hence the RG flow involves the multiple couplings.

\section{On the Cooper pair interpretation}
So far we have not discussed the interpretation of the
fermions developing the superconductivity. In this Section
we make some conjecture concerning their identification
and present the several evidences supporting it. Namely
we shall conjecture that the relevant degrees of freedom
which form the Cooper pairs are the monopoles in the Higgs
branch localized at the nonabelian string.

\subsection{On the monopole interpretation from the knot homologies}
First let us make a few comments concerning the
another appearance of the BA equations for the generalized
Gaudin model. Attempting to get the field theory realization of the
knot homologies the counting of the solutions to the BPS equations
in the proper gauge theory was considered
in \cite{gw}. The problem of counting of the BPS equations was reformulated
in terms of the counting of the solutions to the BA equations
in  the particular integrable system. The mapping of the gauge system
to the BA equations goes as follows.
The inhomogeneities in the BA equation correspond to the t'Hooft lines
or the singular abelian  monopoles. On the other hand the Bethe roots correspond
to the positions of the BPS monopoles at the particular plane. The
BA equations reflect the interaction between the t'Hooft lines and the
BPS monopoles which is S-dual to the process of the W-boson exchange
between two Wilson lines.

The second ingredient of the generalized Gaudin concerns the nontrivial
symmetry breaking in the gauge theory which yields the constant term in the
BA equation. It is this term that gets renormalized when the $\Omega$ deformation
is added. We expect that upon the deformation the theory
enjoys the cyclic RG behavior. In the brane terms the relevant five-dimensional
theory involves the set of M2 branes whose coordinates are partially
fixed by minimizing the superpotential which is generated by the string M2
instanton. The positions of the BPS monopoles are fixed by the BA equations.

The relationship between the knot homologies and the cyclic RG seems
to be not accidental. The point is that cyclic RG can be discovered
in the quantum mechanics of  Calogero -like potential \cite{calogero}
when the Calogero coupling constant is subject of renormalization.
On the other hand the Calogero model at the rational coupling constant is tired
intrinsically to the toric knots.

\subsection{Cyclic RG in Liouville-Toda}
Since there is the AGT relation between the
Nekrasov partition function of conformal SQCD and the 2d conformal
block one could ask about the interpretation of the
Cooper pairs at the Liouville/Toda side. The relation
between the wave functions in the Richardson model and the
conformal blocks in the perturbed WZW model  defined
on the spectral plane of the fermionic model has been analyzed
in \cite{sierraconf}.

Let us briefly comment on  the dictionary between the BSC-like model
and the perturbed conformal theory found in  \cite{sierraconf}.
The Richardson wave function is defined by the number of the
Bethe roots and can be treated as the conformal block in the $\beta-\gamma$
system iin the perturbed conformal theory. The
coupling constant is introduced into the conformal  model via the operator
\beq
V_g= exp(-\frac{i\alpha_0}{g}\oint_C z\partial \phi(z))
\eeq
where $\phi(z)$ is the conventional scalar boson. The Richardson
wave function enters the integral representation of the perturbed
conformal block in the same manner as the Gaudin wave function
enters the integral representation of the Liouville conformal
block (see, for example \cite{teschner}).  The BA equation for the
Richardson wave function corresponds to the saddle point equation in the integral
representation for the conformal block.

The most important
observation from the dictionary obtained in \cite{sierraconf} is that the Cooper pair
operator is attributed to the screening operator and the SL(2)
algebra in the Gaudin model was identified with the algebra of screenings.
Semiclassically these screening operators are attached to the
surface operators hence we have no contradiction with the
conjectured monopole interpretation. The generalization
of the dictionary above  to the RD model should be similar since the
Gaudin model is just substituted by the twisted inhomogeneous XXX spin chain.
We believe that the interpretation of the Cooper pair as screenings works in this case as well.

Since the Nekrasov partition function for  superconformal QCD is related to the
conformal blocks in Liouville and Toda theories is is natural to ask
what is the possible  interpretation of the cyclic RG flows in the Liouville/Toda side.
We restrict ourself by the simple remarks postponing the analysis for the
separate study. The cyclicity could be looked at in the conformal blocks
of the Toda theory corresponding to the conformal SQCD. It would correspond to the
decoupling of the single flavor with the largest mass which  enters the conformal dimensions
of the vertex operators. Simultaneously the gauge coupling constant which corresponds
to the position of the particular vertex operator should be renormalized. One could
expect that the Efimov states in the Liouville/Toda theory could manifest themselves
as the particular resonant states similar to the example elaborated in \cite{ft}.

\subsection{Bion condensates}
If the fermions are identified as the monopoles in the
Higgs phase one could concern on the physical mechanism
providing the formation of the bound states with monopole
charge $Q_M=2$. The possible mechanism providing
the bound states of monopoles has been suggested
in \cite{unsal1,unsal2}. It is based on the consideration of the
SYM theory at $R^3\times S^1$ geometry where the interesting solution
with the several quantum numbers exists. The most relevant solutions
involve both fractional topological and magnetic charges. The compact dimension
provides the breaking of the gauge symmetry and  the finite number
of the vacuum states.

In this geometry there are also so-called KK monopoles which
provide the closeness of the monopole array. The instanton itself
gets interpreted as the bound state of the array of monopoles and KK monopole
with vanishing magnetic and unit topological charge.
The bound state found in \cite{unsal1} involves the monopole
with the fractional topological charge and the KK-antimonopole
with the opposite topological charge. Hence the state does not
have topological charge at all.

The bion condensation has been considered in 3+1 dimensions
however the bulk monopole has the kink counterpart
at the string worldsheet. Hence we could conjecture that
there are the bound states with the kink charge two
similar to the bion condensate in the bulk. The
naive analysis of the zero modes responsible for the
attraction supports this possibility however the
detailed analysis is required.

The RG analysis of the model involving the gas of bions and
electrically charged  W-bosons has been considered in \cite{unsal2}
where the RG flows involves the fugacities for electric and magnetic
components and the coupling constant. The coupled set of the
RG equations has been solved explicitly in the self-dual case and the solution
to the RG equations for the fugacities obtained in  \cite{unsal2}
is identical to the solution for the coupling in the RD model
upon the analytic continuation. The period of the RG in the solution above is fixed
by the RG invariant which has been identified with the product
of the UV values of the electric and magnetic fugacities $y_e\times y_m$.
The similarity between the RG behavior is not accidental
since the mapping of the gauge theory and the perturbed
XY model has been found in \cite{unsal2}.

In the previous sections we have argued that in the deformed
SQCD the period of the cycle is fixed by  graviphoton
field hence one could wonder if the product of fugacities has any
relation with the graviphoton background. The possible answer
could be as follows. The presence of the magnetic and electric
components in the plasma simultaneously implies the possible
angular momentum of the field. 
On the other hand it is possible to identify the analogue of the 
gravimagnetization of the
gauge theory in the RD model. In the gauge theory it is related
to the projection of the angular momentum in the Euclidean 4d space-time \cite{ANGULAR}.
The similar differentiation of the RD Hamiltonian
with respect to $\epsilon$ at $\epsilon=0$ yields 
\beq
\frac{dH_{RD}}{d\epsilon}= \sum_{i<j} (b_ib^{+}_j -b^{+}_jb_i)
\eeq
which is related to the projection of 
angular momentum as well upon use of the commutation relation
in $SL(2,R)$.

Such interpretation suggests the possible place of the full
Efimov tower. As we have mentioned the  new state
which appear(disappear) during the RG cycle corresponds to the
shift in the branch of the logarithm. This is usually
attributed to the additional flux of the global symmetry charge.
Hence the natural conjecture is that the higher Efimov states
are the bound states of the dyons instead of the monopoles.
Such states exist only in some region of the moduli space
decaying at the corresponding curves of the marginal stability.
Note that the relation of the bion ensembles and the perturbed
XY model and interesting interplay of different scales has been recently
discussed in \cite{unsal3}. We plan to discuss the relation between
perturbed XY models  in \cite{unsal3} and in our paper elsewhere.

\section{Discussion}

In this paper we have focused on the two aspects of the relation
between the models of truncated BCS-like superconductivity and
supersymmetric gauge theories
in the graviphoton background. It was demonstrated that the vacuum structure
of the $D=4$ bulk theories and $D=2$ worldsheet theory at the nonabelian
string is inherited also by the third dynamical system - some version of the
superconducting system. The Bethe equations defining the vacuum
structure of the gauge theory yield the spectrum of the excited
states of the Cooper pairs in the superconducting RD model. This can
be considered as an complicated condition of the self-consistency between
the several components of the whole system. The consistency between the
worldvolume theories at D4 and D2 branes with monopole excitation
generalizes the 4d/2d duality for the   nonabelian strings.

The second surprising ingredient of the correspondence concerns
the RG behavior. It was known for a while that  RG behavior
in the bulk and worldsheet theories are consistent. Here we have
the third finite-dimensional RD model
subsystem which  enjoys the peculiar cyclic RG behavior.
This cyclic behavior can be
recognized in the $D=2$ worldsheet theory and $D=4$ bulk theory. In worldsheet theory
the cycle corresponds to the shift of the branch of the solution. This
corresponds to the additional flux along the nonabelian string. In the $D=4$
theory it can be attributed to the change of the branch of the superpotential as well.
In the bulk theory the change of the branch corresponds to the additional  flux
string in the bulk theory which fits with the transformation of the magnetic
string into dyonic one. Let us emphasize that the cyclic RG implies the
possibility of the multiple nonperturbative $\Lambda_i$ scales like
the multiple gaps in the
superconducting model. There is some similarity with the scenario
with the several nonperturbative scales discussed in \cite{ggm1}.

It is natural to ask if there are other possible manifestations of the
cyclic RG behavior in the SUSY gauge theories.
Let us mention two possible candidates. First
note that since   we have observed the  cyclicity in the number
of flavors the
Seiberg duality can be looked at from the new viewpoint. Indeed
generically the Seiberg dual system has the additional mesonic degrees
of freedom and the spectrum is reordered. One could assume that
the "RG time" in this case is
\beq
t_{RG}= log (\frac{N_F}{N_C})
\eeq
and the transition from the initial to the Seiberg  dual system could be treated
as the period of the cyclic RG flow. Naively one needs the second
dimensionless parameter to get the two-coupling flow. In this case
the possible candidate is the ratio of  adjoint and fundamental masses.
Recently the bulk Seiberg duality
has been recognized in the worldsheet theory \cite{seiberg} where
it corresponds to peculiar interchange of the
scale and orientational moduli of the semilocal nonabelian
string. The spectrum
in the dual worldsheet theory has the additional degrees of
freedom as required.

Secondly it was found recently that the spectrum
of the stable BPS particles in the worldsheet theory manifests
the interesting periodicity \cite{bsy}. Namely the curves of the marginal
stability are organized as concentric curves and the
transition from one curve to the next one corresponds to the
shift of the branch of the superpotential.  The curve of the marginal
stability is the very promising place to look at for the whole
Efimov tower. Indeed as we have discussed above the Efimov states
appear when the third degree of freedom is added to the two-body
system near threshold. It is this situation which is realized near
the CMS where the bound state of two BPS constituents
disappears. It seems that phenomena observed in \cite{bsy} could be
the manifestation of the RG cycles and we plan to discuss
this issue elsewhere.

The are a few evident directions of the generalization.
First, it would be interesting to generalize the
analysis above to the asymmetric XXZ and XYZ twisted spin chains
corresponding to the 5d and 6d gauge theories \cite{ggm2}. One could expect
the corresponding superconducting systems for each case where
the anisotropy parameters should be the particular coupling constants.
It would be also interesting to discover the superconducting model
involving two independent  parameters of the $\Omega$ deformation.

It is evident that the investigation of the  cyclic RG flow
in the field theories is
at the very beginning and much more work is required. Among the
most immediate questions a few could be mentioned.

\begin{itemize}

\item  The anomalies can be equally considered as  the IR and UV phenomena
since they have interpretation as the spectrum flow at any scale. This means that
anomalies are a kind of  invariants  of the RG cycle. How this could be formulated
in an invariant manner?

\item What is the holographic image of the cyclic RG flow and the
cycle step?

\item  Is it possible to get the bound state of regulator(UV)  and physical
(IR) degrees of freedom similar to the higher states in the Efimov tower?

\item Is it possible to get the refined partition function involving
the Efimov states or resonances and is there any relation between
such partition functions and the knot invariants? The natural candidate
to look at is the monodromy of the spectrum under the RG cycle. 

\end{itemize}

We hope to investigate these questions elsewhere.

I am grateful to E. Gorsky, N. Nekrasov,  and  A. Vainshtein for the useful discussions.
The work  was partially supported by grants RFBR-12-02-00351 and CRDF-RUP2-2961-MO-09.
The results of the paper were reported at the Conference "Advances in QFT", Dubna
October 4-6.
I am grateful to the Simons Center for Geometry and Physics, KITP at UCSB
during the program " Nonperturbative effects and dualities in QFT and Integrable
Systems" and FTPI at
University of Minnesota where  parts of the
work have been done for the hospitality and support.

\end{document}